\documentclass[conference, 10pt]{IEEEtran}

\IEEEoverridecommandlockouts

\usepackage{graphicx,epstopdf, amsmath,epsfig,subfigure,endnotes,footnote,multicol,sidecap,url,colortbl,booktabs,algorithm,algorithmic,multirow,enumerate}
\usepackage{amsfonts}
\usepackage [english]{babel}

\usepackage{balance}

\usepackage{amsthm}
\usepackage{mathtools}
\usepackage{environ}
\theoremstyle{plain}
\usepackage{comment}

\usepackage{color}
\definecolor{red}{rgb}{1,0,0}

\addto\captionsenglish{\renewcommand{\figurename}{Fig.}}

\begin{document}

\title{Trajectory Synthesis for a UAV Swarm to Perform Resilient Requirement-Aware Surveillance:\\ A Smart Grid-based Study}

\author{
\IEEEauthorblockN{M. Ashiqur Rahman\IEEEauthorrefmark{1}\IEEEauthorrefmark{4}, Rahat Masum\IEEEauthorrefmark{2}\IEEEauthorrefmark{4}, 
Matthew Anderson\IEEEauthorrefmark{3}, and Steven L. Drager\IEEEauthorrefmark{3}{\thanks{\IEEEauthorrefmark{4}Rahman and Masum are the co-first author of this paper.}}}

\IEEEauthorblockA{\IEEEauthorrefmark{1}Department of Electrical and Computer Engineering, Florida International University, Miami, FL, USA\\
\IEEEauthorrefmark{2}Department of Computer Science, Tennessee Tech University, Cookeville, TN, USA\\
\IEEEauthorrefmark{3}Air Force Research Laboratory Information Directorate, Rome, NY,USA\\
Email: marahman@fiu.edu, rmasum42@students.tntech.edu, matthew.anderson.37@us.af.mil, steven.drager@us.af.mil}}


%
%
%
%
%

\maketitle

\begin{abstract}
A smart grid is a widely distributed engineering system with overhead transmission lines. Physical damage to these power lines, from natural calamities or technical failures, will disrupt the functional integrity of the grid. To ensure the continuation of the grid's operational flow when those phenomena happen, the grid operator must immediately take steps to nullify the impacts and repair the problems, even if those occur in hardly-reachable remote areas. Emerging unmanned aerial vehicles (UAVs) show great potential to replace traditional human patrols for regularly monitoring critical situations involving the safety of the grid. The critical lines can be monitored by a fleet of UAVs to ensure a resilient surveillance system. The proposed approach considers the \textit{n}-1 contingency analysis to find the criticality of a transmission line. We propose a formal framework that verifies whether a given set of UAVs (i.e., a UAV swarm) can perform continuous surveillance of the grid satisfying various requirements, particularly the monitoring and resiliency specifications. The verification process ultimately provides a trajectory plan for the UAVs, including the refueling schedules. The resiliency requirement of inspecting a point on a line is expressed in terms of a $k-$property specifying that if $k$ UAVs fail or compromised still there is a UAV to collect the data at the point within on time. We evaluate the proposed framework on synthetic data based on various IEEE test bus systems. 
 \end{abstract}


\section{Introduction}
\label{sec:Intro}

Overhead power lines provide the primary engineering infrastructure within a smart grid, connecting the substations and the generation facilities, to transmit electrical energy along long distances. These transmission lines are distributed from busy cities to remote country areas, often running through coastal ranges, deep forests, long rivers, and mountains.
Natural calamities or technical errors can incur physical damages to the overhead power lines, 
which can hamper the necessary energy transmission, leading the system into an unstable or outage state. The situation worsens when the damage repair is delayed.
Moreover, spatial properties, such as temperature, elongation, and wind induced conductor motion of the transmission lines are important to be regularly taken care of for the optimal infrastructure health~\cite{Khawaja17}. 
An unanticipated natural catastrophe (e.g., wildfire) or extreme weather can deteriorate the condition of one or more weak power lines, which can ultimately cause these lines to break. 
Frequently monitoring the health of the system can minimize this possibility .

Power line surveillance is traditionally event-based, i.e., when the control center detects an outage, technicians perform damage assessment by vehicles. 
%
An alternative, and recently widely used, approach is sending trained inspectors by helicopters to assess the lines for damage using binoculars or cameras~\cite{ma2004aerial}. 
However, these approaches are unsafe, especially during disastrous situations or in the case of remote areas that are hard to reach.
Apart from these drawbacks, the event-based surveillance delays the 
response time during hazardous situations. 
For power-line health maintenance, spatial parameters and electrical properties must be frequently monitored~\cite{Chen15}. 
Hence, {\bf \slshape continuous monitoring} of the critical transmission lines is advantageous, and often a necessity. However, continuous human patrol-based monitoring, even using helicopters, are infeasible due to its high operating cost and potential safety factors. Unmanned aerial vehicles (UAVs) can feasibly replace the human-based patrols~\cite{srinivasan2004airborne, trasvina2017unmanned, li2015uav}. UAVs are flying internet of things (IoT) with network connectivity capabilities.
The emergence of UAVs, and particularly the rapid advancement of their corresponding technology and their increasing cost-efficiency and availability, makes UAV-based surveillance the perfect solution for continuously inspecting overhead transmission lines, even in the event-based scenarios. 

Transmission lines can be put in a critical overload when a line trips or a generation outage happens. The overloaded lines can cause cascading subsequent trips if necessary recovery steps are not taken in time. 
Hence, it is important to analyze the impact of line trips on the system's stability. The contingency analysis, a core component of the energy management system (EMS) in a smart grid, performs this task and selects the operating points (e.g., generation dispatches at different generators) that keep the system in a stable situation even in contingencies, including transmission line or generation source failures. 
Because of the connectivity between the buses and various loads and generations at the buses different transmission lines often impact the system differently. Hence, some transmission lines can be highly critical 
while a few others may not be critical at all. 
Continuous monitoring of the transmission lines should consider their respective {\bf \slshape criticality}. 

\begin{figure*}[t]
\centering
\includegraphics[width=0.8\textwidth]{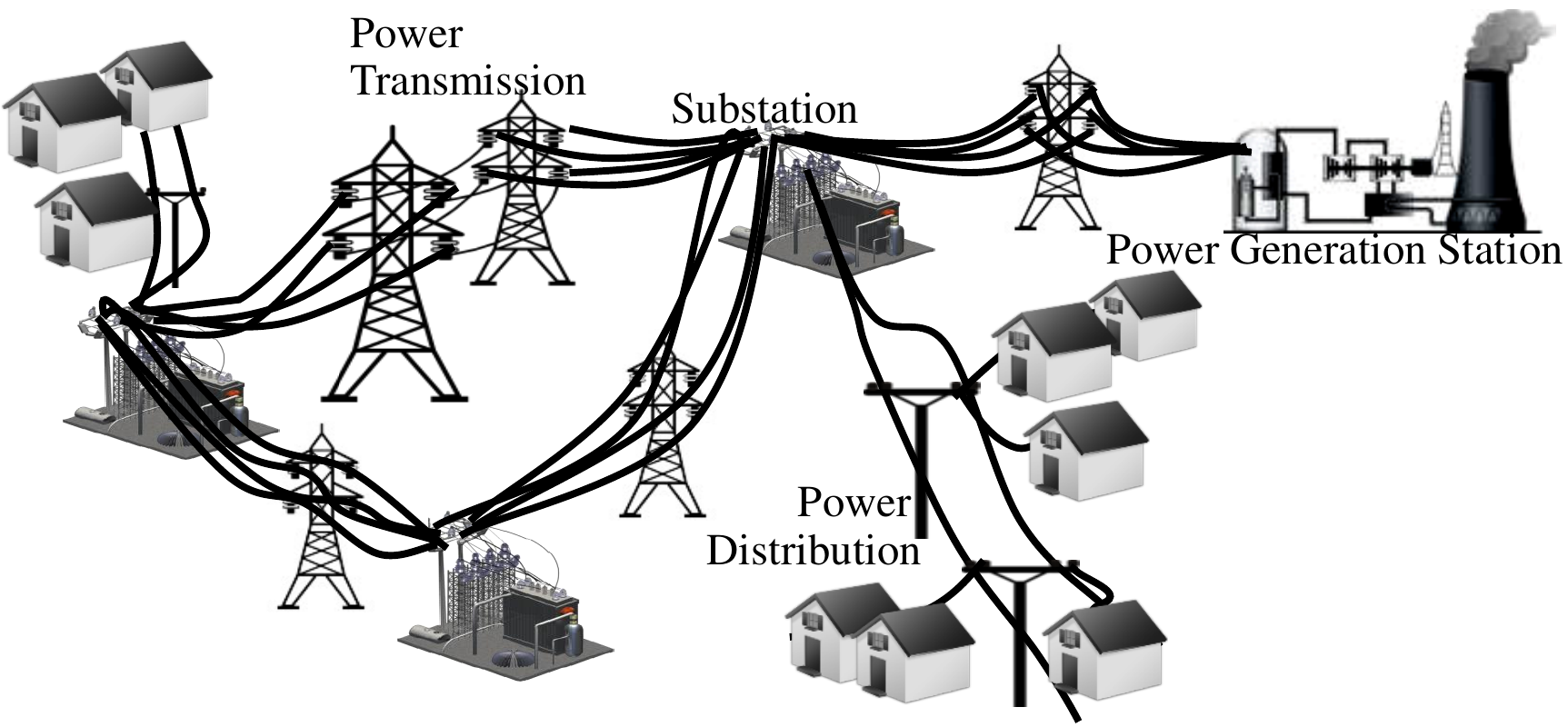}
\caption{Overhead power transmission lines in a smart grid.}
\label{Fig:Lines}
\end{figure*}

Continuous surveillance includes various requirements. While surveillance essentially covers the critical points, the main goal of continuous inspection is maintaining the data freshness (i.e., the subsequent data/image collected at a point on the transmission lines should be within a threshold time frame). 
The surveillance coverage requirement may include all the critical points or some of them that together cover at least a certain portion of the system's overall criticality.
While a UAV can fail or malfunction due to technical errors, it is vulnerable to cyber attacks, especially in an adversarial environment. Therefore, the surveillance \textbf{resiliency} is important. If one or more UAVs fail or are compromised, the properly functioning and uncompromised UAVs may not be able to collect and report data to satisfy a minimal data freshness level. 

Appropriate deployment of a set of UAVs, i.e., a UAV swarm, depends on the feasibility of a trajectory plan for the UAVs that satisfies the surveillance requirement, $k-$resiliency property, and cost-effective fuel usage and refueling making the scenario a hard combinatorial problem. 
This research solves this problem by providing the trajectory plan, along with the refueling schedule, for each UAV, assuming that each UAV is connected to a base control center and deliver captured monitoring data to the base while flying over critical power lines. 
In summary, the major contributions of our work are as follows:
\begin{itemize}
\item We propose a formal model to synthesize a plan for continuous surveillance of overhead transmission lines using a UAV swarm to satisfy the surveillance requirements. 
\item We define the criticality of a transmission line leveraging the contingency analysis. 
We define the resiliency property 
specifying that a data point is under resilient surveillance even if $k$ UAVs are unavailable or compromised. 
\item We implement the proposed formal model using SMT~\cite{Moura09} and demonstrate the solution on a synthetic case study, adapted from an IEEE test bus system~\cite{Testsystems}. 
\end{itemize}
The rest of the paper is organized as follows: In Section~\ref{Sec:Background}, we discuss background of this research and related work. 
We present the proposed formal model for surveillance planning in Section~\ref{Sec:Formal}. We demonstrate the model execution on a case study in Section~\ref{Sec:Case} and present the evaluation results in Section~\ref{Sec:Evaluation}. Finally, Section~\ref{Sec:Conclusion} concludes the paper.

\section{Background}
\label{Sec:Background}

We briefly discuss about the transmission line surveillance and line criticality. Then, we present a related literature review. 

\subsection{Power Transmission Line Surveillance}
\label{SubSec:Background:Surveillance}


The transmission line trips can occur for several reasons, including equipment aging and natural disasters. 
Due to aging, 70\% of the transmission lines have an average age of 25 to 30 years~\cite{campbell2012weather}. Hence, these components need to be inspected regularly in order to maintain uninterrupted power transmission to consumers. 
Severe weather conditions like hurricanes, tornadoes, and wildfires can cause severe destruction of the transmission lines,
including leaning poles, broken wires, tree encroachment over lines, etc. 
These damages affect the electric supply, and the longer the recovery stage, the larger the financial loss.
%
Although natural calamities cannot be controlled in spite of forecasts, damage assessment needs to be addressed as far out and as quickly as possible. 
Incidents like animal attacks, strong winds, and construction work can also trip a power line. 
A regular, continuous monitoring process will help quickly gather information about such incidents. 

In addition to the above scenarios, adversaries can attack the control center by making injurious operating decisions, which can overload the transmission lines, ultimately leading to the destruction of equipment. A frequent or continuous surveillance may detect the potential failure and save the line from being tripped, and thereby the system from an outage.
%

\subsection{Transmission Line's Criticality Analysis}
\label{SubSec:Background:Criticality}

EMS is the core component of the bulk energy management in a smart grid and consists of several interdependent computational modules~\cite{Abur04, Wood13}. 
EMS executes these modules based on the measurement data received from the field devices and accordingly control the grid (e.g., generation set-points) for operational security and economic efficiency. Contingency analysis (CA) is one of the core EMS modules.
The goal of CA is to operate the power system securely by analyzing the system subject to a contingency (e.g., transmission line outage) and determine the set-points that will allow system operation without violation of constraints. Typically, $n - 1$ contingency analysis is performed where  $n$ is the total number of nodes (either transmission lines or generation sources) and one node failure is considered~\cite{Abur04, Wood13}. The analysis considers the failure of each transmission line and checks its impact on the system, e.g., whether the rest of the transmission lines become overloaded. In this respect, some lines are more critical than the rest as one of  failures often have more negative impacts on the system than some others.

We usually compute the criticality of a transmission line using line outage distribution factors (LODFs)~\cite{Abur04, Wood13}. However, these factors depend on the topology, electrical properties of the lines, and loads and generation dispatches at the buses. A change in the load or the generation dispatch impacts the criticality of the system. The calculation of LODFs is briefly presented in Appendix~\ref{App:LODF} for the interested readers.
All the outage cases are ranked according to a performance index ($\mathcal{P}$) calculation.
%
Let $\mathbb{L}$ be the set of all lines in the system, 
$P_l$ be the power flow through line $l$, and $\bar{P}_l$ be the flow capacity of the line.
If $L_{l, \hat{l}}$ defines the LODF for line $l$ after an outage of line $\hat{l}$, the change in the power flow ($\Delta P_l$) on line $l$ due to this failure 
can be found as:  $\Delta P_{l, \hat{l}} = L_{l, \hat{l}} \times P_l$.
If the ultimate power flow, $\hat{P}_{l, \hat{l}}$, will be $\hat{P}_{l, \hat{l}} = P_l + \Delta P_{l, \hat{l}}$.
Now, if $n$ is a suitable index, for a contingency, e.g., the outage of line $l'$, the simplest form of the index, i.e., $\mathcal{P}_{l'}$, will be as follows~\cite{Wood13}: 
$\displaystyle\mathcal{P}_{\hat{l}} = \sum_{l \in \mathbb{L}}^{}{\left({\hat{P}_{l, \hat{l}}}/{\bar{P}_l} \right )^{2n}}$.

A larger $\mathcal{P}$ for a line shows that it has a higher criticality than that of a line with a smaller $\mathcal{P}$. In this way, $\mathcal{P}$ values help rank the transmission lines per their critical sensitivities. 
There are improved ways of calculating $\mathcal{P}$ for better understanding of the critical contingencies~\cite{Ranking}. 
\subsection{Related Work}
\label{SubSec:Background:Related}

We discuss the existing literature related to this research in different categories as follows.

\vspace{3pt}
\noindent\textbf{UAV-Based Surveillance Technique.}
Srinivasan et al. presented the idea of video surveillance using cameras and sensors leveraging UAVs in a project with the Florida Department of Transportation~\cite{srinivasan2004airborne}. The video images contain traffic information on the roads and are transmitted by UAVs using the microwave IP network. 
Latchman et al. proposed that UAVs, equipped with a GPS location set and surveillance capabilities, can take direction from the ground control station and determine necessary altitude calculations for surveillance using airborne sensors~\cite{latchman2005airborne}. 
Jaimes et al. provided some aspects of the real-time image recognition task based on the videos sent from the preassigned GPS coordinate-programmed UAVs~\cite{jaimes2008approach}. 
UAVs were used in a project by Moreno et al. to monitor marine environments in the Mexican seashore using cheap sensors and less power-consuming buoys~\cite{trasvina2017unmanned}. 

\vspace{3pt}
\noindent\textbf{Power Line Surveillance.}
%
%
Traditionally, electric utilities send technicians by vehicles to the potential damaged areas for inspecting the towers. 
Ma et al. discussed an alternative approach, in which trained inspectors are sent by helicopters to inspect lines using binoculars or cameras and record data to a log book for further analysis~\cite{ma2004aerial}. However, this approach 
still unsafe in extreme weather conditions and can be still inefficient for remote, unpassable areas. 
Several latter works discussed the use of UAV-based monitoring instead of traditional human patrol methods. 
A project named Hydro-Quebec LineScout Technology
applied remotely-controlled mobile robots/UAVs to perform basic power line inspection and maintenance tasks~\cite{montambault2007inspection, montambault2009remote}. 
Li et al. presented a knowledge-based power line detection method from the captured images so that UAVs can be utilized for surveillance and inspection systems~\cite{li2008knowledge, li2010towards}. 
%
%
Pagnano et al. proposed an automated surveillance mechanism for the transmission lines, leveraging UAVs/robots that rolled on the wires~\cite{pagnano2013roadmap}. This real-time inspection methodology uses the image and signal data processing, allowing the detection of faults or abnormalities on the lines. 
A UAV-based system for high voltage power line inspection is presented in~\cite{luque2014power}, where for real-time error reporting, quadcopters were equipped with color cameras controlled from the ground control station. 

\vspace{3pt}
\noindent\textbf{Optimal Surveillance Design.}
Semsch et al. proposed a two-stage mechanism: (i) constructing a covering point set and (ii) exploring trajectory based on UAV motion through the points, which can control flights for autonomous multi-UAVs to allow maximum surveillance~\cite{semsch2009autonomous}. The mechanism looks for a set of paths so that, in case of obstacles, a minimum area will be uncovered, giving the adversaries fewer opportunities for exploitation.
Lim et al. targeted the challenges associated with using UAVs to scan power lines from a distance and send the damage data to the control center~\cite{lim2016multi}. The authors presented a solution
to minimize overall inspection time and cost by prepositioning the UAVs optimally. 
%
However, the approach is event-based, allowing for monitoring disaster situations with the prepositioned UAVs. 
Deng et al. proposed a multi-platform cooperative UAV system as well as a multi-model communication system for power line inspections in China~\cite{deng2014unmanned}. The authors considered several design challenges, particularly the delay between image capturing. 

\vspace{3pt}
\noindent\textbf{UAV's Surveillance Security.}
The UAV monitoring system can succumb to cyber attacks that allows sensitive data to be collected by the adversaries. Birnbaum et al. presented a real-time behavioral monitoring procedure that can convert a flight plan to behavioral profiling~\cite{birnbaum2015unmanned}. 
%
Abbaspour et al. demonstrated the safety-critical issues and corresponding detection mechanisms for the UAV-based surveillance with respect to several faults and sensor-spoofing attacks~\cite{abbaspour2016detection}. 

\begin{figure*}[t]
\centering
\includegraphics[width=0.8\textwidth]{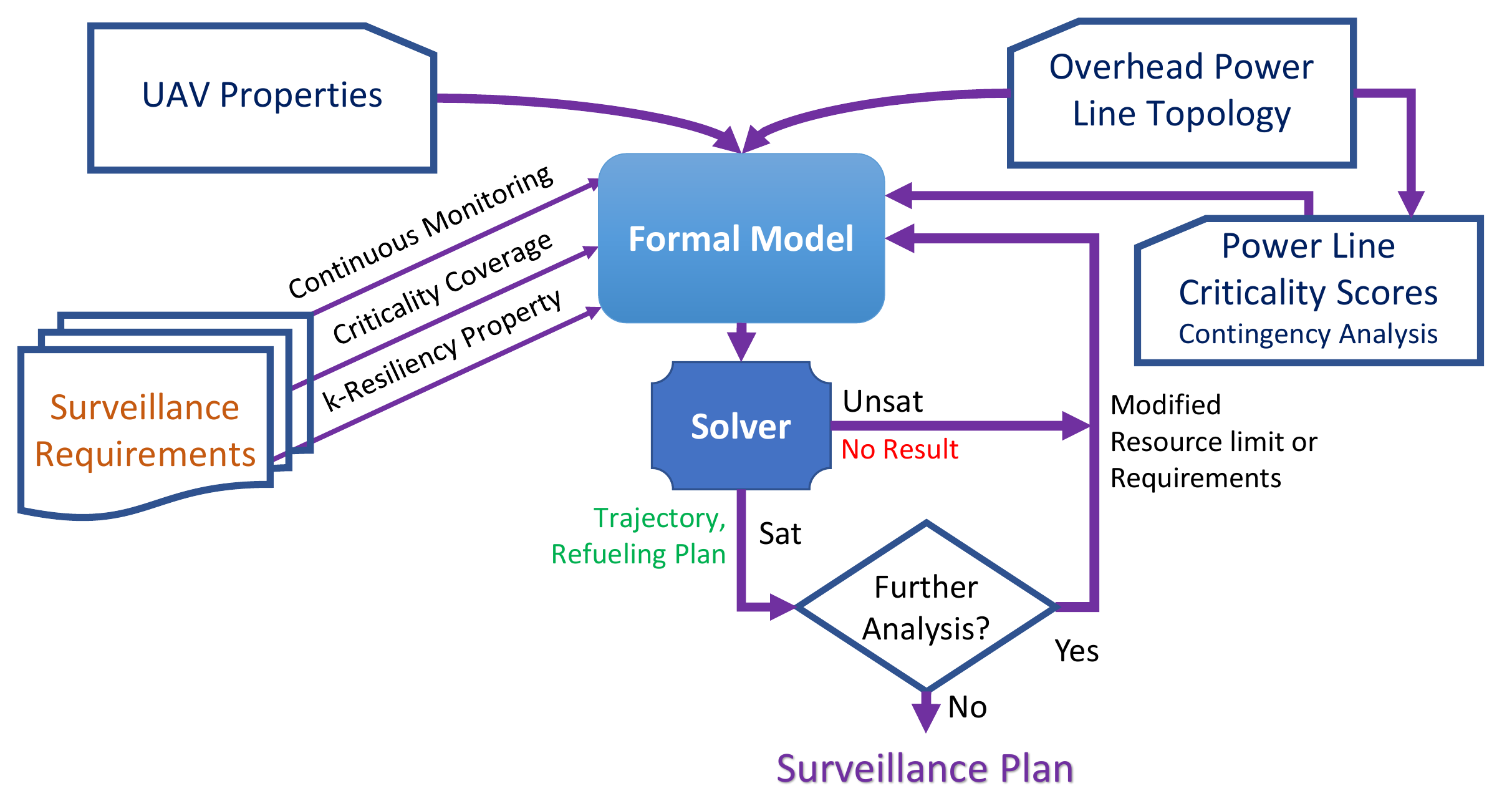}
\caption{The solution approach for resilient surveillance planning and analysis.}
\label{Fig:Architecture}
\end{figure*}

\subsection{Contributions}
While the existing literature presented techniques for UAV-based surveillance of power transmission lines, an automated and efficient trajectory planning mechanism for continuous surveillance is important, especially when different lines can have different criticality levels, and these levels frequently change with the change in the topology, loads, or generation dispatches. 
Since some UAVs can fail or be compromised, resiliency of the surveillance against such events is crucial. However, automated synthesis of a resilient surveillance plan, i.e., trajectories of a fleet of UAVs, including their refueling schedules,  under a resource limitation is a combinatorially hard problem. This research provides a solution to this complex problem by formally modeling the \textbf{continuous} and \textbf{resilient} surveillance properties as a constraint satisfaction problem. 
%
%
%
Fig.~\ref{Fig:Architecture} shows the proposed resilient surveillance planning framework. 
%
A formal model is developed and solved to provide a routing strategy, along with a refueling plan, for the UAV fleet satisfying the continuous and resilient surveillance requirements and corresponding constraints. 

\section{Resilient Surveillance Model}
\label{Sec:Formal}

%
%


In this section, we present the system model and define the resilient surveillance accordingly.

\subsection{System Model}

The UAVs do the continuous surveillance of the transmission system by flying over (within a safe distance) the power lines~\cite{Zhang17}. We assume that each UAV's surveillance trajectory follows the bus topology. 
A line is divided into several segments, with each segment (link) connecting two points. We assume a fixed segment length, 
and hence, the number of segments or points on a transmission line represents its physical length. The surveillance is modeled through visiting these points.
Each UAV, identified using an ID, possesses a set of properties like its average speed, fuel capacity, initial fuel, and starting position on the line topology. In this modeling, we also assume that all UAVs maintain the same speed on an average throughout the surveillance period irrespective of the altitude differences or necessary turns between the points. However, different UAVs may need different fuel consumption to maintain the speed.

Let $\mathbb{P}$ denote the set of points on the network and $\mathbb{U}$ be the set of UAVs to perform the surveillance. Moreover, 
$\mathbb{P}_l$ ($\mathbb{P}_l \subseteq \mathbb{P}$) is the set of points on line $l$ and $\mathit{Seg}_{p, p'}$ denotes if points $p$ and $p'$ are connected.
Since the UAVs have the same speed, the same amount of time will be taken to cover each segment. 
We consider this required time as one time unit. The surveillance will be modeled for a period of (analysis) time, say $S$ time units, where $s$ ($1 \le s \le S$) will identify a particular time step. 
Each UAV $u$ has a fuel capacity ($\mathit{FuelCap}_{u}$) and it starts surveillance with an initial fuel ($\mathit{InitFuel}_{u}$) from a specific point ($\mathit{InitPoint}_{u}$). Let $\mathit{Fuel}_{u, s}$ be the remaining fuel of UAV $u$ at step $s$, $\mathit{FFuel}_u$ be the fuel required to fly a segment (i.e., fuel consumption at each time unit), and $\mathit{HFuel}_u$ be the fuel required for hovering during a time step. We use $\mathit{TB}_p$ to denote the time steps to go to the base or the (closest) refueling station from point $p$ for refueling. 

\subsection{UAV Trajectory Model}

Let $\mathit{Visit}_{u,p,s}$ denote whether UAV $u$ is visiting point $p$ at time $s$ ($s > 1$). UAV $u$ can visit point $p$ at $s~(> 1)$ in two cases. In the first case, the UAV is already there (i.e., if it is hovering/loitering), and it has sufficient fuel for hovering during the time step. In the second case, the UAV is at point $p'$ that is connected to point $p$ ($\mathit{Seg}_{p', p}$ is true), and it has sufficient fuel to fly there. In both of the cases, the sufficient fuel requirement also includes the fuel needed to fly to the refueling station from the visited point ($p$). This constraint ensures that no UAV will be out of fuel and be stranded. We formalize the first case ($\mathit{Hover}_{u, p, s}$) as follows:
\begin{equation*}
\label{Eq:Hover}
\begin{split}
\mathit{Hover}_{u, p, s} \rightarrow~ &\mathit{Visit}_{u, p, s - 1} \wedge \\
& (\mathit{Fuel}_{u,s} = \mathit{Fuel}_{u,s - 1} - \mathit{HFuel}_{u}) \wedge \\
& (\mathit{Fuel}_{u, s} \ge \mathit{FFuel} \times \mathit{TB}_p) 
\end{split}
\end{equation*}


In the second case ($\mathit{Fly}_{u, p, s}$), the fuel consumption cost depends on the climbing angle of the segment ($\mathit{Seg}_{p', p}$). This impact of climbing angle (upward or downward) on the fuel consumption is abstracted using a ratio ($\mathit{CRatio}_{p', p}$) of the required cost to fly that segment over that of flying the same length of a horizontal segment. The following equation presents the corresponding formalization:
\begin{equation*}
\label{Eq:Fly}
\begin{split}
\mathit{Fly}_{u, p, s} \rightarrow \bigvee_{p' \in \mathbb{P}, p' \neq p} & \mathit{Seg}_{p', p} \wedge \mathit{Visit}_{u,p',s-1} ~\wedge \\
(\mathit{Fuel}_{u, s} & = \mathit{Fuel}_{u, s - 1} - \mathit{FFuel} \times \mathit{CRatio}_{p', p}) ~\wedge\\
(\mathit{Fuel}_{u, s} & \ge \mathit{FFuel} \times \mathit{TB}_p)
\end{split}
\end{equation*}

Therefore, visiting a point $p$ at time $s$ by UAV $u$ ($\mathit{Visit}_{u,p,s}$) is defined as follows:
\begin{equation}
\label{Eq:Visit}
\mathit{Visit}_{u, p, s} ~\rightarrow \mathit{Hover}_{u, p, s} \vee \mathit{Fly}_{u, p, s}
\end{equation}
%
%
We assume that if no UAV partially covers a segment partially, i.e., if the UAV starts flying from a point over a segment, it reaches the end point of the segment. The same is true about hovering at a point during a time step.
The initial location (at $s = 1$) of each UAV $u$ is identified at some point $p$ on the topology according to its initial (current/given) placement.

A point is connected with two or multiple points. At a particular time step, one UAV can only choose one segment. The constraint is formalized as follows:
\begin{equation}
\forall_{u,p,s}\mathit{Visit}_{u,p,s} ~\rightarrow \bigwedge_{p' \in \mathbb{P},~p \neq p'} \neg \mathit{Visit}_{u,p',s}
\end{equation}

$\mathit{Visited}_{p,s}$ denotes whether point $p$ has been visited by any UAV at step $s$. Hence:
\begin{equation}
\mathit{Visited}_{p,s} \rightarrow \bigvee_{u \in \mathbb{U}} \mathit{Visit}_{u,p,s}
\end{equation}


\subsection{Refueling Model}


The refueling of a UAV is modeled by abstracting its path from a point $p$ to the base or a refueling station and subsequently returning to a point $p'$. For simplicity of presenting the refueling model, we assume only one refueling station. The path distance from a point to the refueling center is often more than one time step ($\mathit{TB}_p$). We define $\mathit{ToRefuel}_{u, p, s}$ to denote that UAV $u$ is moving to the station from point $p$ at time $s$ for refueling, $\mathit{RefuelTo}_{u, p, s}$ to represent the return of UAV $u$ after the refueling to point $p$ at time $s$, and $\mathit{Refuel}_{u, s}$ to specify UAV $u$ is refueling at the station at time $s$.
If $\mathit{ToRefuel}_{u, p, s}$ is true, then the following equation holds:
\begin{equation}
\begin{split}
\mathit{ToRefuel}_{u, p, s} \rightarrow & \mathit{Visit}_{u, p, s} \wedge \mathit{Refuel}_{u, s + \mathit{TB}_p} \\
& ~\wedge \left(\bigwedge_{p \in \mathbb{P}} \bigwedge_{s < s' \le s + \mathit{TB}_p} \neg \mathit{Visit}_{u, p, s'} \right)
\end{split}
\end{equation}
We also need to ensure that $\mathit{Refuel}_{u, s}$ is true only if there is a valid $\mathit{ToRefuel}_{u, p, s -  \mathit{TB}_p}$ for some point $p$. 
Similarly, if $\mathit{Refuel}_{u, s}$ is true, then the following equation must hold:
\begin{equation}
\begin{split}
\mathit{Refuel}_{u, s} \rightarrow \bigvee_{p \in \mathbb{P}} & \mathit{Visit}_{u, p, s + \mathit{TB}_p} \wedge \mathit{RefuelTo}_{u, p, s + \mathit{TB}_p} \\
&  \wedge \left(\bigwedge_{p \in \mathbb{P}} ~\bigwedge_{s < s' < s + \mathit{TB}_p} \neg \mathit{Visit}_{u, p, s'} \right)
\end{split}
\end{equation}
It is also ensured that $\mathit{RefuelTo}_{u, p, s}$ is true only if there is a valid $\mathit{Refuel}_{u, s -  \mathit{TB}_p}$, and this $p$ is the only return point for this particular refueling. 

We assume that a UAV refuels to its capacity. The remaining fuel after refueling, more appropriately after returning to point $p$ for resuming the surveillance task, is computed as follows:
\begin{equation}
\mathit{RefuelTo}_{u, p, s} \rightarrow \mathit{Fuel}_{u,s} = \mathit{FuelCap}_{u} - \mathit{FFuel}_{u} \times \mathit{TB}_p
\end{equation}

\subsection{Continuous Surveillance}

The continuous surveillance for a point requires that it is always visited at least once within a (given) threshold period of time ($\mathit{TC}$). In other words, each pair of two consecutive visits to this point is done within $\mathit{TC}$. Let $\mathit{Surveilled}_p$ denote whether point $p$ is continuously surveilled in $S$.
%
In this case, if point $p$ is visited at time $s$, the next visit to this point needs to be at some time $s'$ within $\mathit{TC}$:
\begin{equation}
\label{Eq:Surveilled}
\begin{split}
\mathit{Surveilled}_p \rightarrow  \bigwedge_{1 \leq s \leq (S - \mathit{TC})} & \mathit{Visited}_{p,s}  \\
& \rightarrow \bigvee_{s < s' \leq (s + \mathit{TC})} \mathit{Visited}_{p,s'}
\end{split}
\end{equation}
%
%
To make the continuous surveillance true from the beginning (to initiate the above equation to act for all points), $\mathit{Surveilled}_p$ must also ensure that starting from the beginning within the threshold time there is at least one visit to point $p$: 
\begin{equation*}
\mathit{Surveilled}_p \rightarrow ~\bigvee_{1 \leq s \leq \mathit{TC}} \mathit{Visited}_{p,s}
\end{equation*}

\subsection{Resilient Surveillance}

A point is under $k-$resilient surveillance if it is visited by $k + 1$ (different) UAVs within a (given) threshold time ($\mathit{TR}$) throughout the surveillance period. We define that point $p$ is under resilient surveillance ($\mathit{ResVisited}_{p,s}$) at time $s$ (for the time period $\mathit{TR}$) if the following equation holds:
\begin{equation*}
\begin{split}
& \mathit{ResVisited}_{p,s} \rightarrow \mathit{Visited}_{p,s} \wedge \\
& ~~~~(\sum_{u \in \mathbb{U}} \mathit{VisitDuring}_{u, p, s, \mathit{TR}} \ge (k + 1)) \vee  ((s + \mathit{TR}) \le S))
\end{split}
\end{equation*}
Here, $\mathit{VisitDuring}_{u,p,s,\mathit{TR}}$ denotes that whether point $p$ is visited by $u$ during the period from $s$ to $\mathit{TR}$. That is:
\begin{equation*}
\mathit{VisitDuring}_{u,p,s,\mathit{TR}} \rightarrow ((s + \mathit{TR}) \leq S) \wedge
\bigvee_{s \leq s' \leq s + \mathit{TR}} \mathit{Visit}_{u,p,s'} 
\end{equation*}

Then, the continuous monitoring requirement is ensured for resilient surveillance by the following constraint:
\begin{equation}
\label{Eq:ResSurveilled}
\begin{split}
\mathit{ResSurveilled}_p ~\rightarrow & ~\mathit{Surveilled}_p ~\wedge \\
\bigwedge_{1 \leq s \leq (S - \mathit{TR})} & \mathit{ResVisited}_{p, s} \rightarrow \\
& \bigvee_{s < s' \leq (s + \mathit{TR})} \mathit{ResVisited}_{p, s'}
\end{split}
\end{equation}


\subsection{Criticality Coverage Requirements}

The objective of the surveillance is to continuously monitor the transmission system (e.g., power lines, generators, or other physical components) such that the surveilled points cover at least a threshold part of the overall criticality. We define the criticality coverage score as the criticality of the points under continuous surveillance over the total criticality of the system. We consider that a point on a line will have the same criticality weight, $\mathit{PC}_p$), as that of the line, $\mathit{LC}_l$, which can be equal to $\mathcal{P}_l$~(Section~\ref{SubSec:Background:Criticality}) or a scaled value~(for further details, see Appendix~\ref{App:Weight}):
\begin{equation*}
\forall_{p \in \mathbb{P}_l} \mathit{PC}_p = \mathit{LC}_l
\end{equation*}
However, for a point at which two or more lines connected (i.e., at a substation), the criticality weight will be the maximum of the corresponding lines' weights. 
%

If $\mathit{CS}$ denotes the minimum requirement of the criticality coverage score for the continuous surveillance, then the following should be satisfied:
\begin{equation*}
\displaystyle \frac {\sum_{p \in \mathbb{P}} \mathit{Surveilled}_p \times \mathit{PC}_{p}} { \sum_{p \in \mathbb{P}} \mathit{PC}_{p}} \ge \mathit{CS}
\end{equation*}
For an arithmetic operation on boolean parameters, we assume boolean ``true'' and ``false'' as integer 1 and 0, respectively. 

The resilient (continuous) surveillance requirement is often different than that of the continuous surveillance because the former considers a contingency or attack scenario. If $\mathit{RCS}$ denotes the required criticality score under the resilient surveillance, then we have the following constraint:
\begin{equation*}
\displaystyle \frac {\sum_{p \in \mathbb{P}} \mathit{ResSurveilled}_p \times \mathit{PC}_{p}} { \sum_{p \in \mathbb{P}} \mathit{PC}_{p}} \ge \mathit{RCS}
\end{equation*}

\subsection{Repetition of Surveillance Plan}

Since the surveillance is continuous and the model considers a particular surveillance period, the operator may repeatedly follow the same trajectory plan or execute the model for continuous surveillance periods considering the last execution result as the input for the next run. This is because the model cannot be solved efficiently for a long surveillance period as the number of clauses grows rapidly with the time steps. 

In the first case, when the same trajectory plan will be executed constantly, we can achieve this requirement by ensuring a couple of constraints at the end of the surveillance period (i.e., at time step $S$).
First, each UAV $u$ must return to its starting position ($\mathit{InitPoint}_{u}$). Second, the remaining fuel of the UAV at time $S$ needs to be equal to or greater than its initial stored fuel. The following equation formalizes these two constraints:
\begin{equation*}
\displaystyle \mathit{Visit}_{u,\mathit{InitPoint}_{u},S} ~\wedge~ (\mathit{Fuel}_{u, S} \ge \mathit{InitFuel}_u)
\end{equation*}

Lastly, to ensure the continuous surveillance between the last visit to a point in one cycle and the first visit to the point in the next cycle, we redefine Equation~(\ref{Eq:Surveilled}) as follows:
\begin{equation*}
\begin{split}
\mathit{Surveilled}_p \rightarrow  \bigwedge_{1 \leq s \leq S} & \mathit{Visited}_{p,s}  \\
&\rightarrow \bigvee_{s < s' \leq (s + \mathit{TC})} \mathit{Visited}_{p, (s' \% S)} 
\end{split}
\end{equation*}
Similarly, Equation~(\ref{Eq:ResSurveilled}) will be updated.

%

\begin{figure}[t]
\centering
\includegraphics[width=\columnwidth]{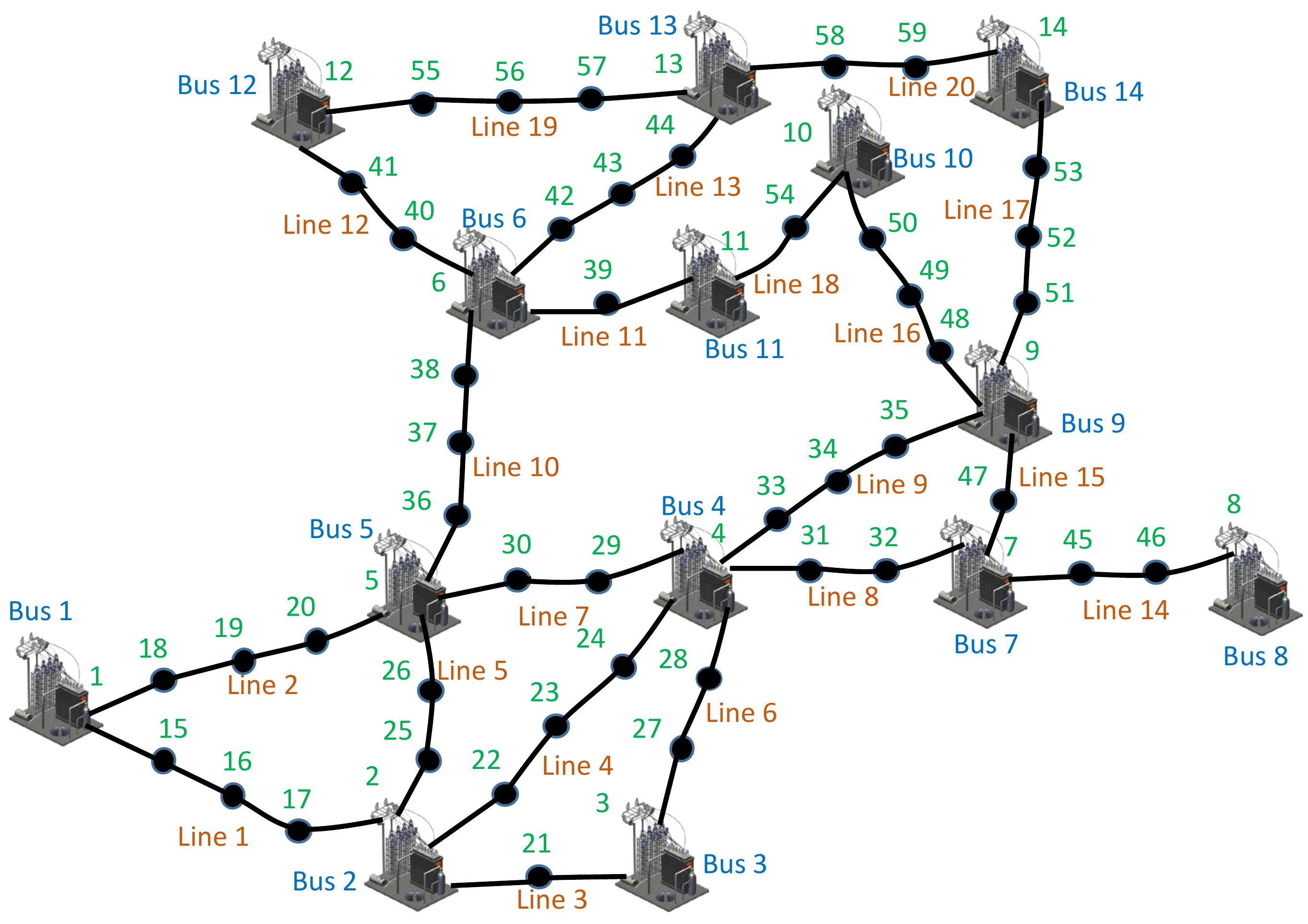}
\vspace{-9pt}
\caption{The example overhead transmission line infrastructure.} 
\label{Fig:Case}
\vspace{-9pt}
\end{figure}

\begin{table}[ht]
\caption{Case Study Input (Partial)}
\label{Tab:Case}
\vspace{-9pt}
\scriptsize
\begin{center}
\begin{tabular}{p{0.95\columnwidth}}
\hline
\vspace{0.01in}
\# Number of Buses, Lines, Points, and Segments, Segment Length in Time Units, Number of UAVs, Surveillance Period\\
14 20 59 65 1 5 91\\


\vspace{0.01in}
\# Load Information (Bus No, Load)\\
4 93.0\\
5 66.0\\
7 66.0\\
9 91.0\\
10 93.0\\
\ldots \ldots \ldots\\

\vspace{0.01in}
\# Generation Information(Bus No, Generation)\\
1 121.0\\
2 18.0\\
3 57.0\\
\ldots \ldots \ldots\\

\vspace{0.01in}
\# Transmission Line Info (From-Bus, To-Bus, Reactance)\\
1 2 0.05917\\
1 5 0.22304\\
2 3 0.19797\\
2 4 0.17632\\
2 5 0.17388\\
3 4 0.17103\\
4 5 0.04211\\
\ldots \ldots \ldots\\

\vspace{0.01in}
\# Maximum Criticality (PI Score) Distance\\
15
\%\\

\vspace{0.01in}
\# Line Point Set\\
1 15 16 17 2 \\
1 18 19 20 5 \\
2 22 23 24 4 \\
2 21 3 \\
2 25 26 5 \\
\ldots \ldots \ldots\\

\vspace{0.01in}
\# Segments/Links (End Points, Fuel Cost Ratio)\\
1 15 1.0 \\
15 16 0.95 \\
16 17 1.0 \\
17 2 1.12 \\
1 18 \\
\ldots \ldots \ldots\\

\vspace{0.01in}
\# UAV Properties (Initial Point, Stored Fuel, Fuel Capacity (Watt), Mileage (Fuel/Step), Hovering Cost (Fuel/Step))\\
10 1200 1500 15 3\\
5 600 1200 12 3\\
14 300 1500 15 6\\
30 1050 1500 12 3\\
\ldots \ldots \ldots\\

\vspace{0.01in}
\# Threshold Time between Two Consecutive Visits to a Point\\
25\\

\vspace{0.01in}
\# Resiliency Requirements (k, Threshold Time)\\
2 45\\

\vspace{0.01in}
\# Minimum Criticality Scores under Continuous Surveillance and Resilient Surveillance\\
80 50
\vspace{0.01in}
\\\hline
\end{tabular}
\end{center}
\normalsize
\end{table}

\section{Implementation and A Case Study}
\label{Sec:Case}

We briefly discuss the implementation of the model and present a synthetic case study. 

\subsection{Implementation}

We use the SMT logic~\cite{Moura09} to encode the formalization presented in the previous section. 
%
The encoded model is solved using Z3, an efficient SMT solver~\cite{Z3}. 
The solution to the model gives a result primarily as \emph{sat} or \emph{unsat}. In the case of \emph{sat}, 
we receive the detailed trajectory planning that can successfully perform the surveillance task, satisfying the constraints. More specifically, the terms $\mathit{Visit}_{u, p, s}$, $\mathit{ToRefuel}_{u, p, s}$, $\mathit{Refuel}_{u, s}$, $\mathit{RefuelTo}_{u, p, s}$, and $\mathit{Fuel}_{u, s}$ provide the trajectory paths and refueling plans, including remaining fuels. 
The developed program reads necessary inputs (i.e., data about the bus topology, the transmission line infrastructure, the UAV fleet, and surveillance requirements) from a text file. The required outputs are printed on different text files.

\subsection{Case Study}

We consider a synthetic overhead power line infrastructure based on the IEEE 14-bus test system~\cite{Testsystems}. There are 14 buses and 20 lines in this test system. The power line infrastructure is shown in Figure~\ref{Fig:Case}, where each transmission line is divided into multiple segments according to its assumed length. There are 65 equal-length (1 mile) segments, each connecting two points on the transmission lines. The points are numbered from 1 to 59. There is a fleet of 5 UAVs to perform the surveillance. The surveillance period is 91 time units/steps. As we assumed in the modeling, each segment is covered in a time step, although the fuel consumption/cost of covering a particular segment depends on the type (mileage property) of the UAV and the climbing angle for this segment.
The partial input file corresponding to this case study is presented in Table~\ref{Tab:Case}.
According to the load and generation information of the buses, there are 5 generation buses (i.e., buses 1, 2, 3, 6, and 8). 
The transmission line information includes the end buses and the impedance (reactance) for each line. 

Overhead transmission line infrastructure information includes the set of points that constitutes each line. For instance, line 1 is constituted of points 1, 15, 16, 17, and 2. The segment or link information more specifically provides information about the segment: the end points and the fuel cost ratio to fly from the first point to the second point in one time step. The ratio is taken over the cost of flying the same distance horizontally (climbing angle is zero). For example, the fuel cost required to fly the segment from point 1 to point 15 is normal, while the flying cost from point 15 to point 16 is 5\% less. The cost ratio is inverted if the flying direction is the opposite. The UAV information consists of a set of properties for each UAV, which includes its starting position, initial stored fuel, fuel capacity, and fuel costs per step for flying (when the climbing angle is zero) and hovering/loitering. It is worth mentioning that these property values are synthetic and driven from practical sources, particularly considering HyDrone UAVs~\cite{Jameson09, HyDroneMMC, Skyfront}.

As the surveillance requirements, we consider (i) continuous surveillance threshold as 25 time steps, 
(ii) 2-resilient surveillance and corresponding threshold as 45 time steps (i.e., a point under resilient surveillance must be visited by $k$ + 1 (3) different UAVs in 45 units), and (iii) the minimum scores under continuous surveillance and resilient surveillance respectively as 80\% and 50\%.

\begin{figure*}[t]
    \begin{center}
	\subfigure[]{
            \label{Graph_UAV_Criticality}
            \includegraphics[width=0.4\textwidth]{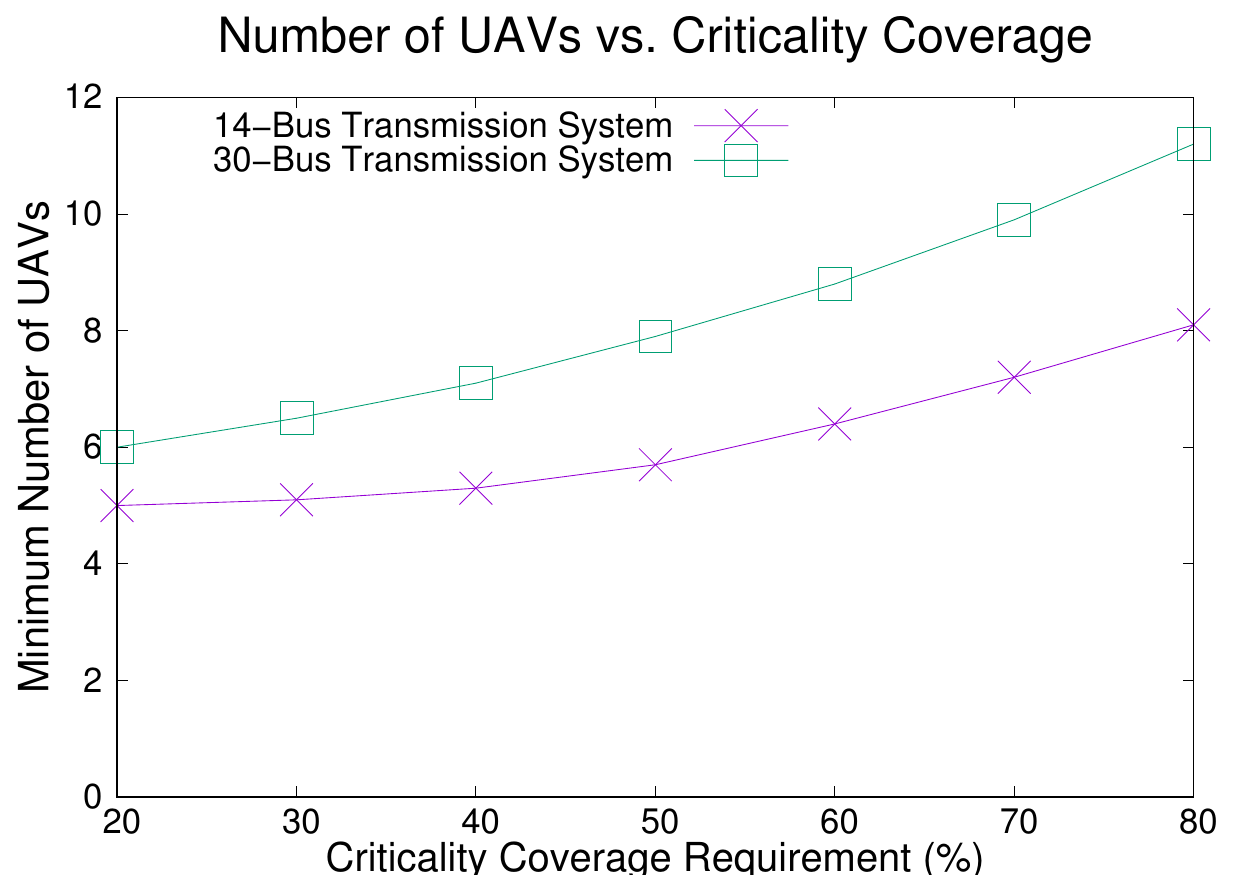}
        }\hspace{12pt}            
     \subfigure[]{
            \label{Graph_UAV_CriticalityDistribution}
            \includegraphics[width=0.4\textwidth]{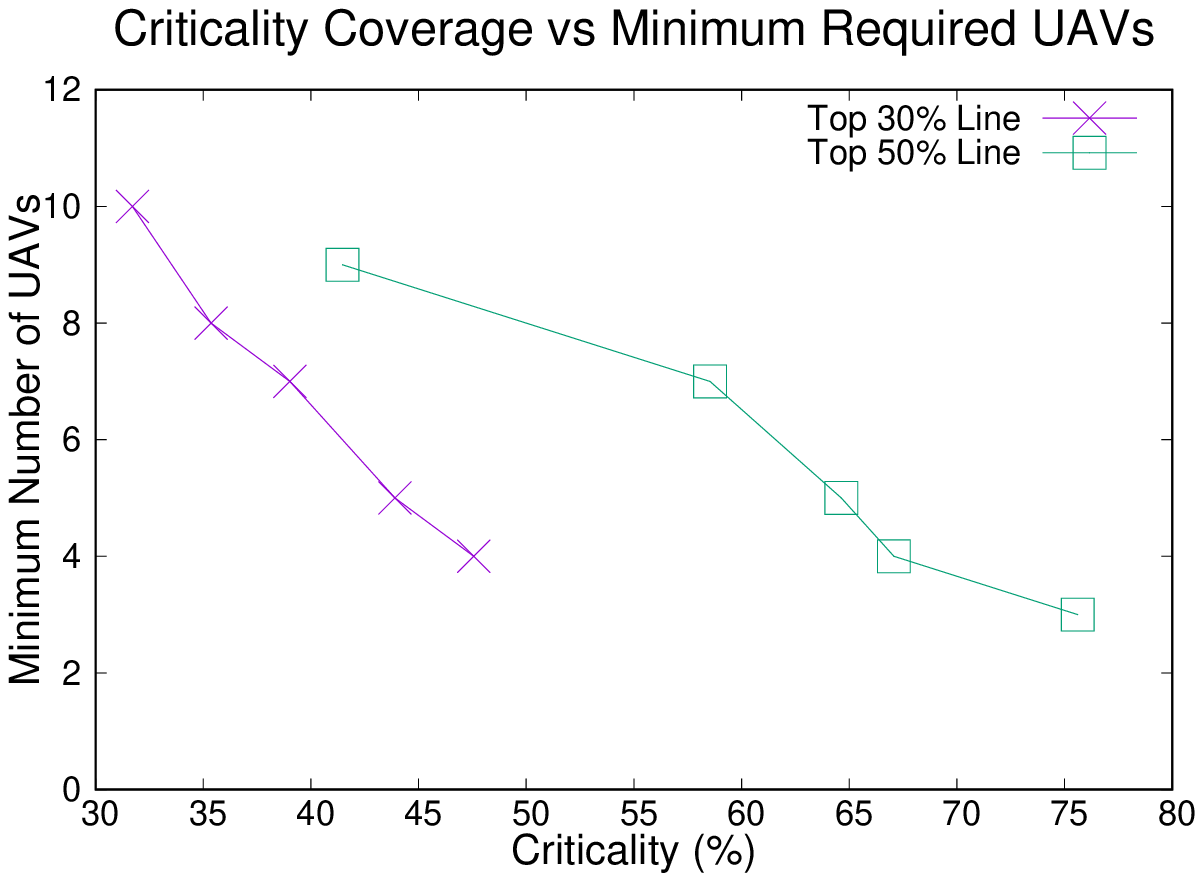}
        }                   
    \end{center}
    \caption{The minimum number of UAVs to perform the continuous surveillance depends of (a) the criticality coverage requirement 
    and (b) the distribution of criticality among the transmission lines.}
\end{figure*}

\begin{figure*}[t]
    \begin{center}
	\subfigure[]{
            \label{Graph_Criticality_Bus}
            \includegraphics[width=0.4\textwidth]{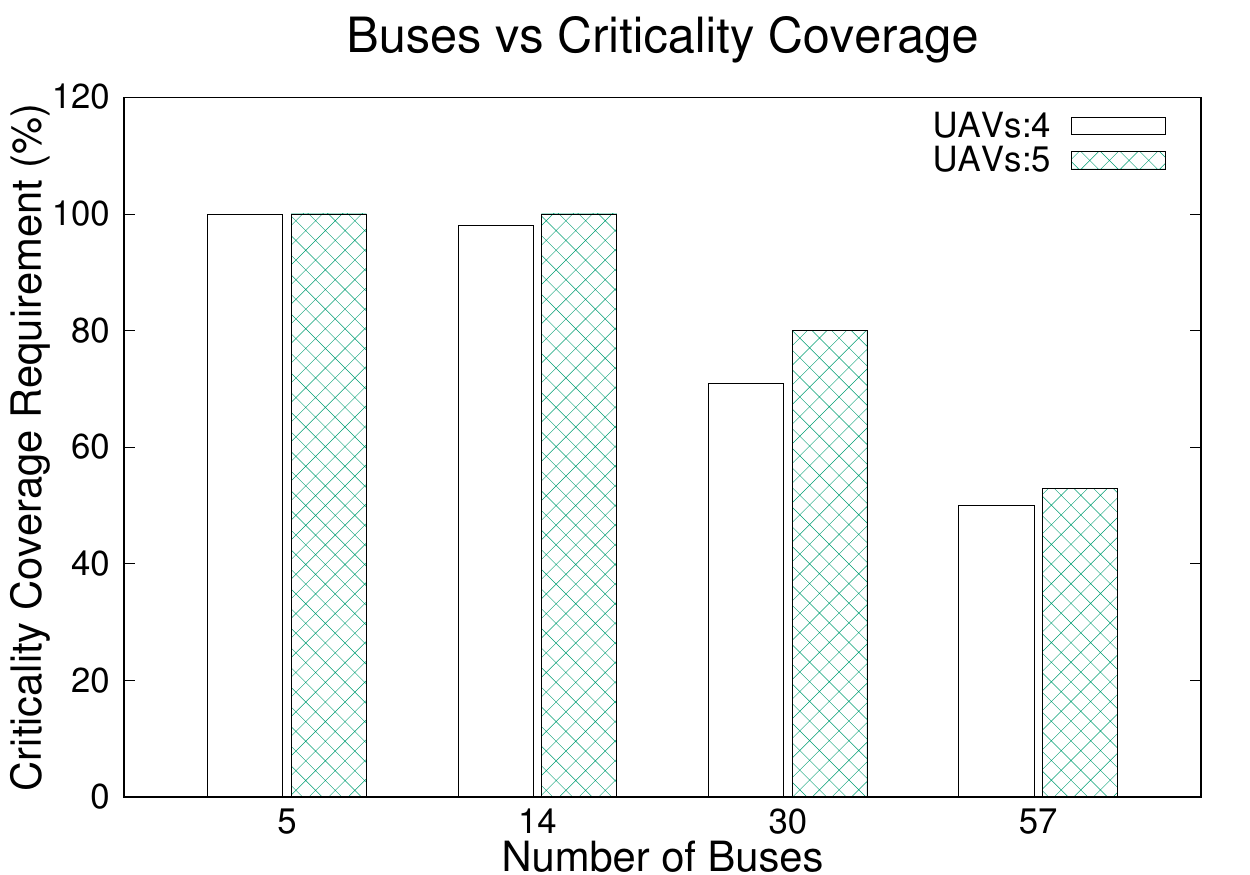}
        }\hspace{12pt}
	\subfigure[]{
            \label{Graph_UAV_Bus}
            \includegraphics[width=0.4\textwidth]{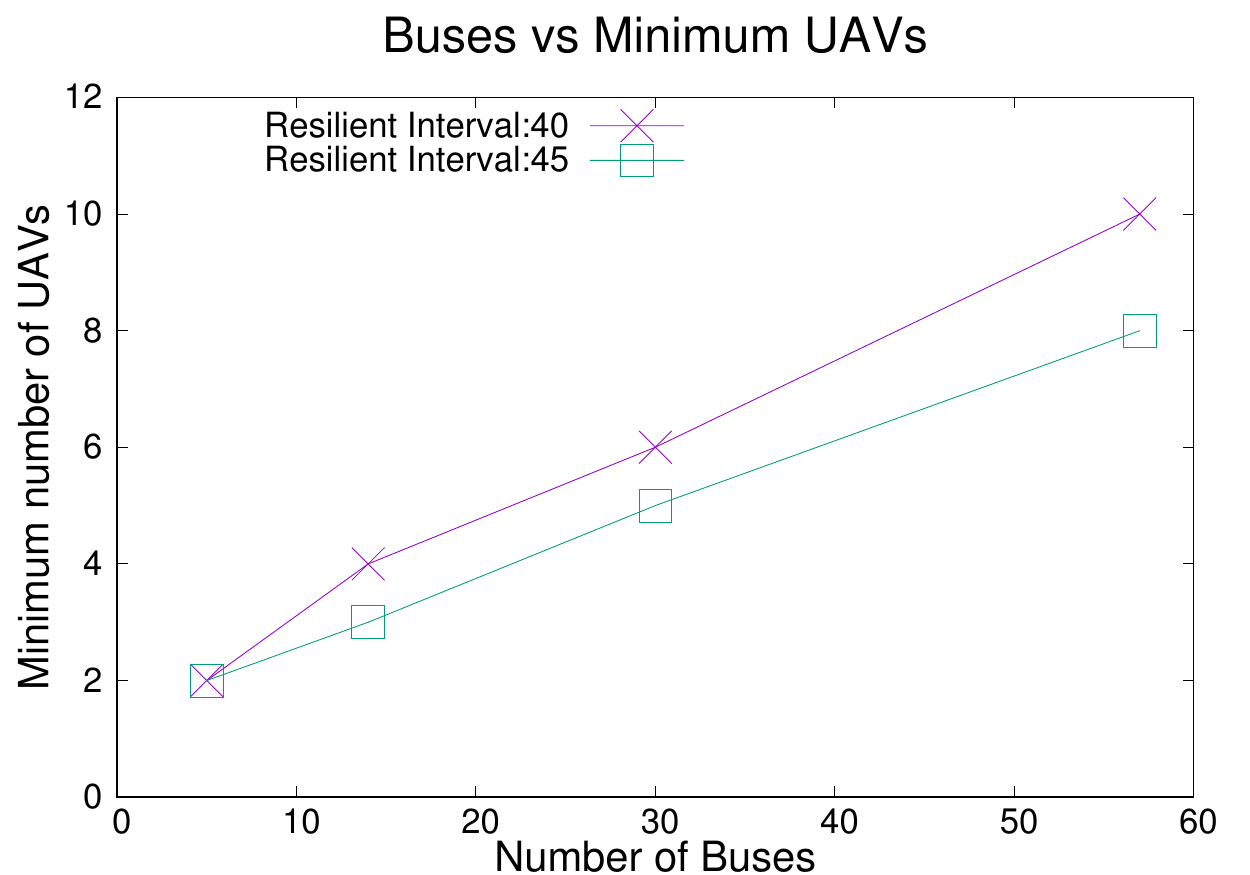}
        }
    \end{center}
    \caption{Impact of the grid size (i.e., the number of buses) on (a) the minimum number of UAVs required for the surveillance and (b) the maximum criticality achieved with a fixed number of UAVs.}
\end{figure*}

The solution to the corresponding formal model provides a \emph{sat} result and provides the trajectory plan, including the refueling schedule. The result shows three criticality levels. According to the trajectory plan, 45 points are under continuous surveillance, covering around 81\% of criticality, while 28 points are under resilient surveillance, covering over 50\% of criticality. During the continuous surveillance, all the high (level 3) criticality points (13 points), three-quarters of the medium (level 2) critical points (17 points), and  65\% of the low critical points are under coverage. The 2-resilient surveillance covers around 62\% of the high critical points (8 points), 48\% of the medium critical points (11 points), and 40\% of the low critical points. Due to the distribution of the higher critical lines in the topology and a limited number of UAVs, many less-critical points are also covered compared to high-critical points. 
If we consider the UAV visits to a particular point, e.g., point 3, the result shows that the point is visited at time step 7 by UAV 1 and so on as follows ([at, by]): [7, 1], [14, 4], [38, 2], [50, 3], [64, 1], and [74, 1]. These visits satisfy not only the continuous surveillance requirement but also the resilient surveillance condition, ensuring the visits of three (k = 2) UAVs followed (and including) from each visit within the threshold time. If we consider point 2, we find that it is only continuously surveilled ([5, 1], [8, 2], [25, 2], [27, 2], [37, 3], [48, 3], [56, 5], [81, 1], and [88, 2]). The point is visited quite a few times but not by a required number of different UAVs within the threshold time. The output also includes the refueling schedule for the UAVs. According to this schedule, for example, UAV 2 goes for refuels at point 9 between times 46 and 54 while UAV 3 refuels at point 58 between times 16 and 24.

\section{Evaluation}
\label{Sec:Evaluation}

We evaluate the proposed surveillance plan synthesis model to analyze surveillance characteristics as well as its scalability.

\subsection{Methodology}

We analyze the characteristics of surveillance by evaluating the minimum number of UAVs to perform surveillance satisfying the criticality coverage requirement and the vice versa. 
The evaluation is performed on different synthetic grid infrastructures, driven from various IEEE test bus systems~\cite{Testsystems}. We consider the system size as the number of buses. The scalability of solving the proposed model is evaluated in terms of the execution time by varying different surveillance requirements.
%
We run our experiments on an Intel Core i7 machine with $16$ GB memory. 


\subsection{Evaluation Results: Characteristic Analysis}

\begin{figure*}[t]
    \begin{center}
	\subfigure[]{
            \label{Graph_Time_Criticality}
            \includegraphics[width=0.3\textwidth]{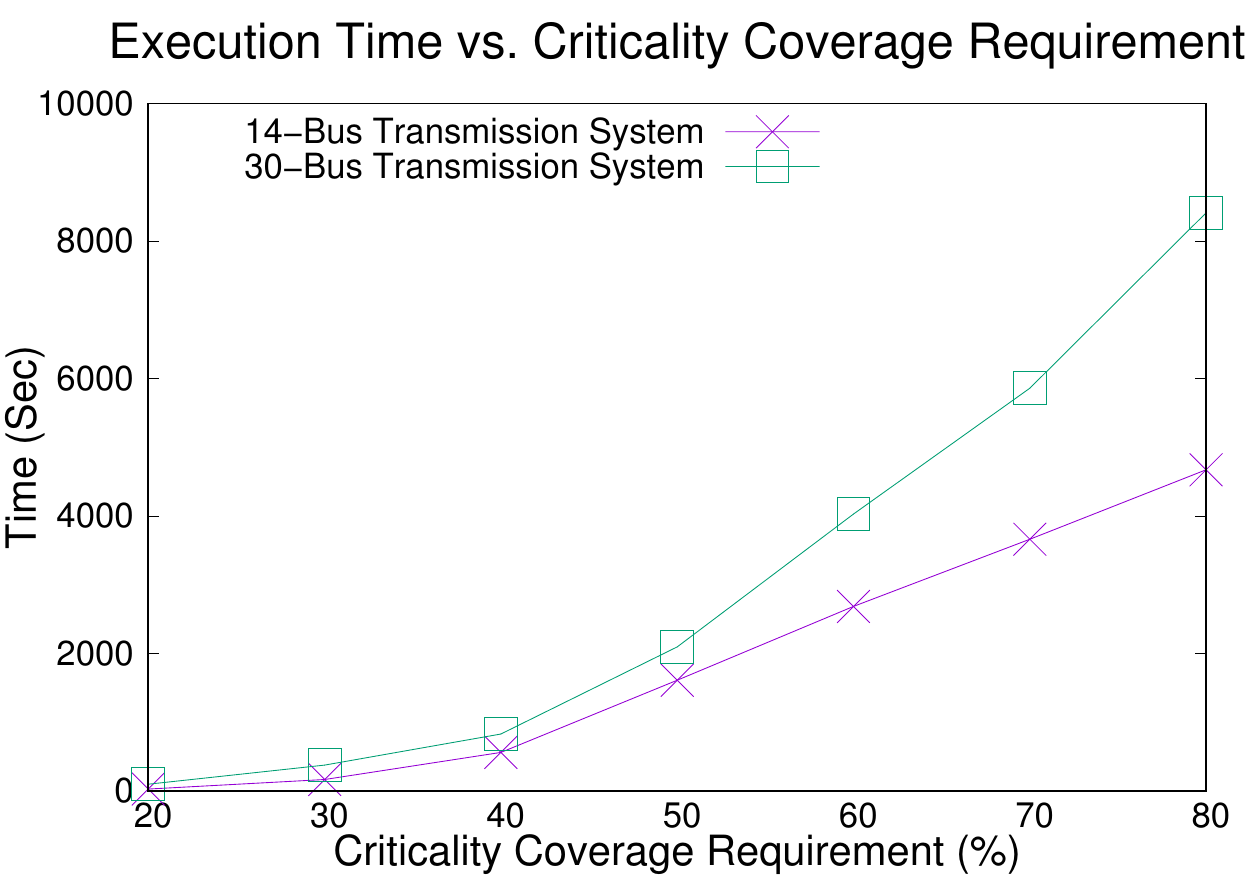}
        }\hspace{3pt}
	\subfigure[]{
            \label{Graph_Time_UAV}
            \includegraphics[width=0.3\textwidth]{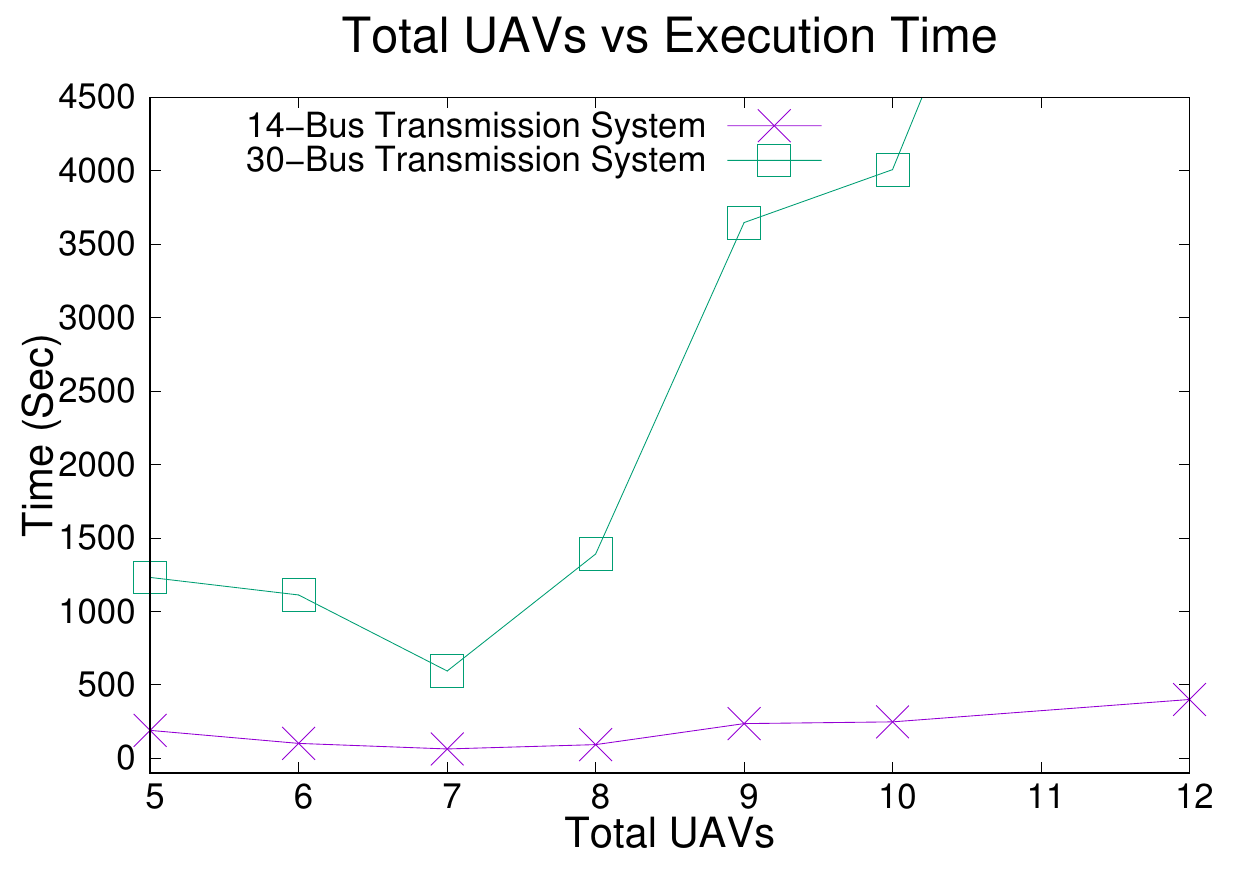}
        }\hspace{3pt}                
	\subfigure[]{
            \label{Graph_Time_Kvisit}
            \includegraphics[width=0.3\textwidth]{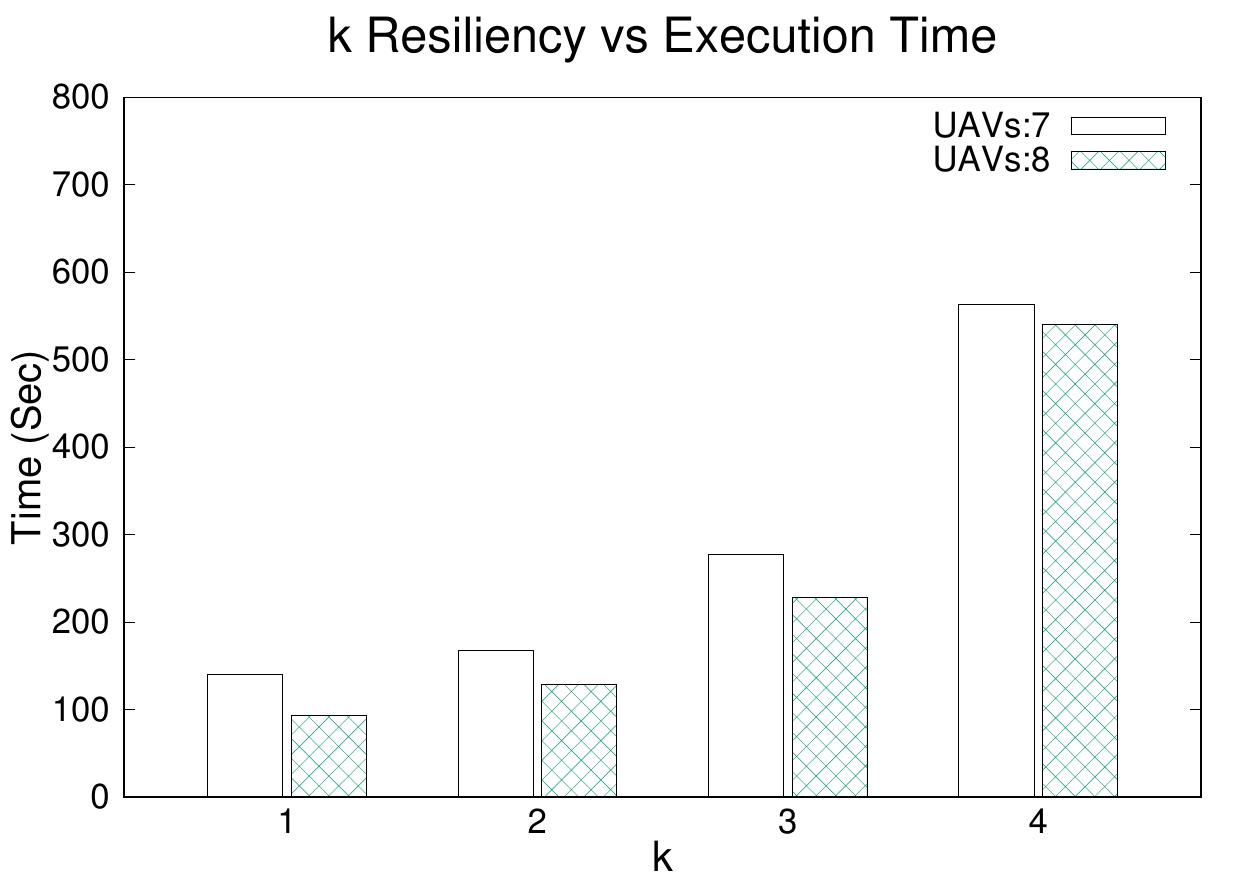}
        }        
%
	\subfigure[]{
            \label{Graph_Time_VisitInterval}
            \includegraphics[width=0.3\textwidth]{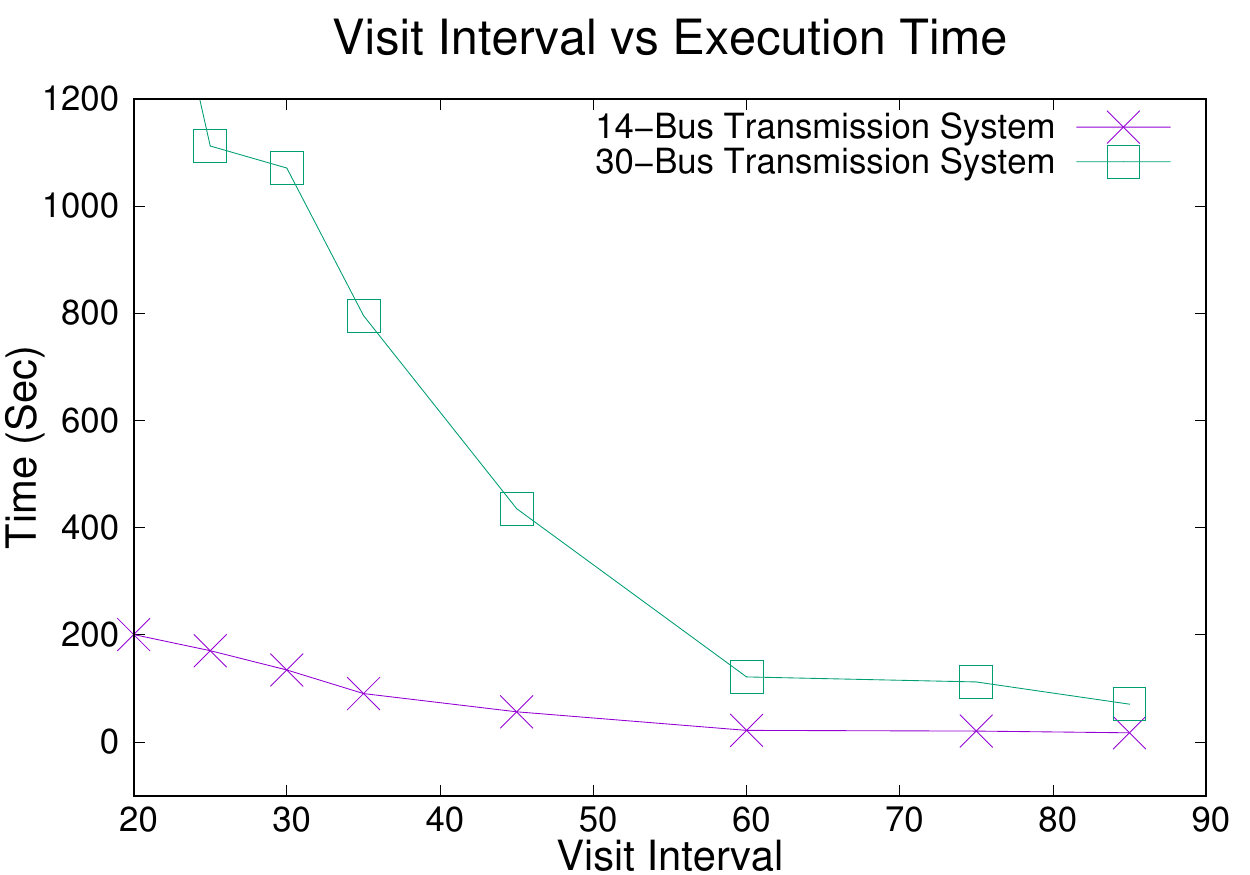}
        }\hspace{3pt}
	\subfigure[]{
            \label{Graph_Time_ResInterval}
            \includegraphics[width=0.3\textwidth]{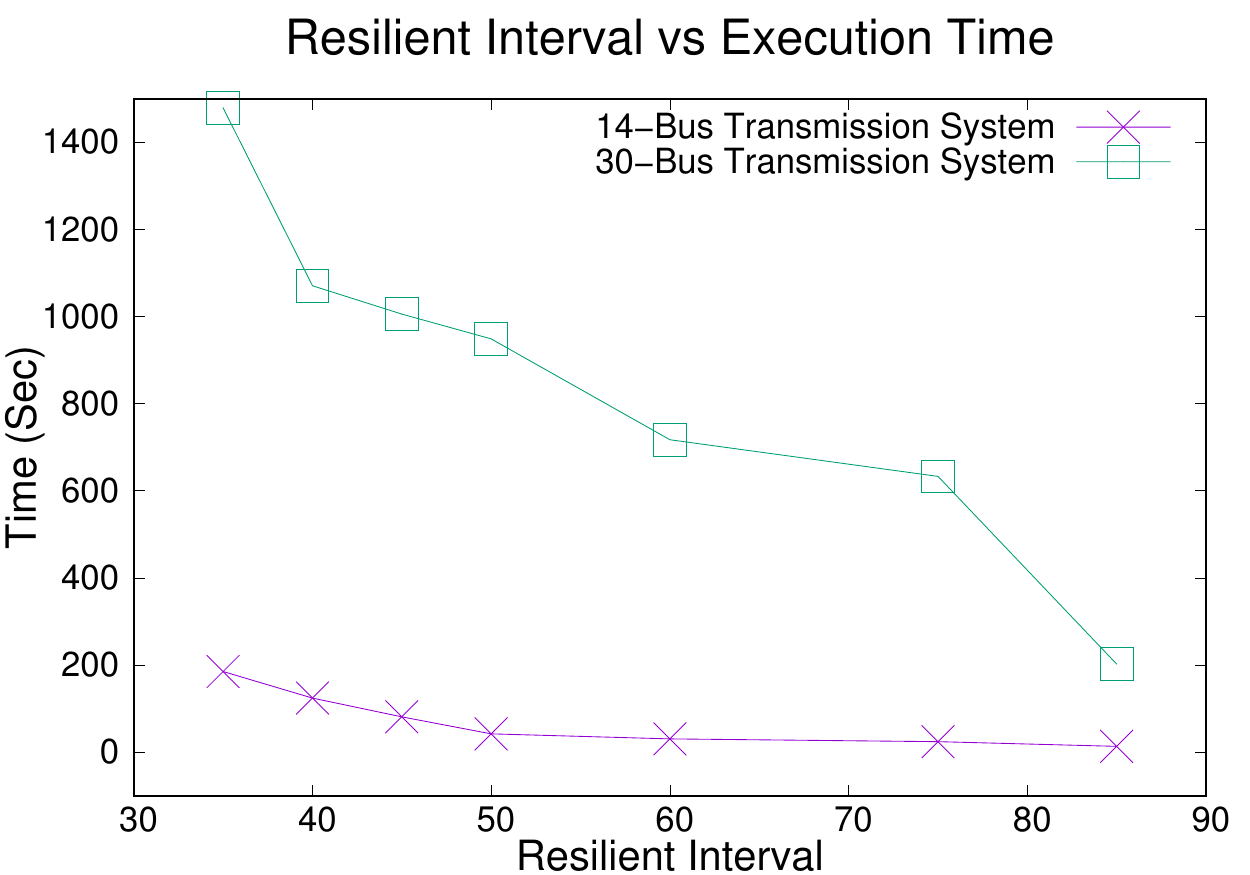}
        }\hspace{3pt}                
	\subfigure[]{
            \label{Graph_Time_Period}
            \includegraphics[width=0.3\textwidth]{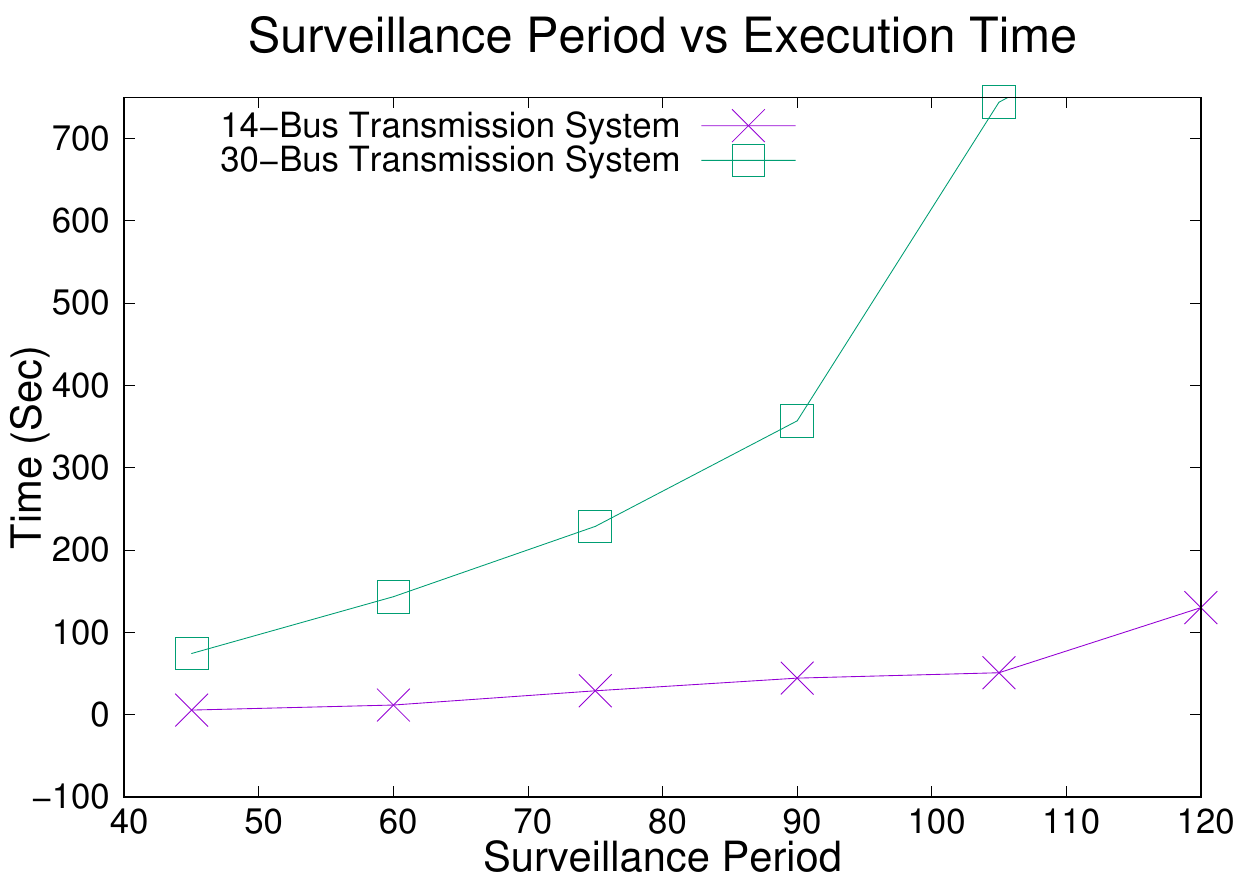}
        }                
    \end{center}
    \caption{Impact of (a) the criticality coverage requirement for continuous surveillance, (b) the size of the UAV fleet (i.e., the number of UAVs), and (c) the selection of $k$ for resilient surveillance, (c) the time interval for the continuous surveillance, (b) the time interval for the resilient surveillance, and (d) the surveillance period on the formal model execution time.}
\end{figure*}

\noindent\textbf{Impact of the Criticality Coverage Requirement on the Number of UAVs for Surveillance.} The synthesis of the surveillance plan, i.e., the trajectory of the UAVs and their refueling schedules, depends on the surveillance requirements, i.e., the continuous surveillance (data freshness) time threshold, resiliency specification and corresponding surveillance time threshold, and the criticality coverage requirements, along with the problem size. The number of UAVs required to perform the surveillance depends on satisfying all these properties. The tighter the constraints, the more UAVs are required to synthesize the surveillance plan. 
We analyze the impact of the criticality coverage requirement by the surveilling UAVs in this respect. The analysis result is shown in Fig.~\ref{Graph_UAV_Criticality}. 
The graph shows that the number of minimally required UAVs increases with the increase in the coverage requirement. This is because to cover higher criticality, a larger area (i.e., a larger set of points) typically needs to be under surveillance. 
%
%
With the problem size (e.g., the number of buses), this number increases further. As the figure shows, the minimum number of UAVs required for surveillance in the case of the 30-bus system is larger than that of the 14-bus system.

\noindent\textbf{Impact of the Criticality Distribution over Lines.} The required number of UAVs to perform the surveillance operation depends on the scenario about how the critical weights are distributed over the transmission lines. Fig.~\ref{Graph_UAV_CriticalityDistribution} shows this impact. 
In this evaluation, we consider two scenarios as follows. We order the transmission lines according to their criticality weights and take two sets of lines, top 30\% and 50\% lines, as the two cases. We consider the same total criticality for these cases. The criticality coverage requirement for the continuous surveillance is 70\%. 
As the graph for the top 30\% critical lines shows, the more criticality this set of lines hold in total, the lesser number of UAVs the surveillance job requires. It is because the UAVs need to consider a smaller set of lines to meet the criticality coverage requirement for the overall transmission line topology. We see the same behavior for the case of the top 50\% lines.

\noindent\textbf{Impact of the Grid Size on the Criticality Surveillance.} We analyze the impact of the grid size (i.e., the number of UAVs) on the maximum criticality coverage by the surveilling UAVs. The analysis result for two different numbers of UAVs is presented in Fig.~\ref{Graph_Criticality_Bus}. 
The result demonstrates that, for a specific number of UAVs, the criticality coverage is limited and with the increased size of the transmission system (i.e., the number of buses) the criticality coverage reduces. The is because a larger system has a larger set of surveillance points, often more critical points, and these points are widely distributed in the broader infrastructure.
%
Therefore, for an increased system size, a particular set of UAVs cannot but surveilled a reduced part of the system's criticality while, with the number of UAVs, the criticality coverage increases (e.g., 4 UAVs vs. 5 UAVs in the figure).
%

\noindent\textbf{Impact of the Grid Size on the Number of UAVs.} 
We analyze the impact of the grid size (i.e., the number of buses) on the number of UAVs required to cover a particular criticality coverage. We consider 60\% and 30\% as the criticality score coverage requirements for continuous surveillance and resilient ($k$ = 1) surveillance, respectively. The analysis result for two different resilient surveillance intervals is presented in Fig.~\ref{Graph_UAV_Bus}. In both cases, the continuous surveillance interval is 25 time steps.
The graphs in the figure specify that a larger grid requires a higher number of UAVs to cover the required criticality score because the surveillance points in a larger system are often more widely distributed. 
%
We can also observe in the figure the impact of resiliency surveillance interval on the number of UAVs. As the figure shows, a larger interval requirement (i.e., 45 steps) often needs a smaller number of UAVs than a smaller one (i.e., 40 steps). This is because a larger interval allows a longer time frame to visit a point $k~+~1$ (2) times, which often reduces the number of UAVs required to achieve the criticality surveillance score. 


\subsection{Evaluation Results: Scalability Analysis}

\noindent\textbf{Impact of the Criticality Coverage Requirement.} Fig.~\ref{Graph_Time_Criticality} shows the execution/solving time of the proposed formal model with respect to the criticality coverage requirement for 14-bus and 30-bus transmission systems. As shown by the graphs, with the increase in the coverage requirement, the time to solve the model grows. The higher is the requirement, the larger the area that needs to be covered, which increases the execution time. When the requirement is close to the maximum possible coverage for a set of UAVs, the search space increases rapidly, which increases the execution time. If a larger set of UAVs is used for the surveillance, the size of the model (i.e., the number of variables and assertions/clauses) also expands. A larger model requires an increased, often exponentially high, solving time. 
A large grid is usually divided into multiple sub-grids (regional grids) to manage the system in a decentralized fashion. For a smaller grid, our proposed formal model can efficiently, even with our limited computing capability, synthesize the trajectory plan for required surveillance.

\noindent\textbf{Impact of the number of UAVs.} 
We evaluate the impact of the number of UAVs on the solving time for a particular problem size and a  surveillance requirement.
Fig.~\ref{Graph_Time_UAV} presents the result. We can observe that initially with  the increase in the number of UAVs the execution time reduces. After a point, the execution time grows rapidly. This is because, initially an increased number of UAVs eases the searching for a solution providing more options. However, as the number of UAVs grows, the number of clauses in the model increases superlinearly, which ultimately increases the execution time. 

\noindent\textbf{Impact of the Resiliency Requirement ($k$).} We analyze the impact of the $k$, the resilient surveillance requirement on the 14-bus system. We consider two fleets of UAVs on various $k$ values for a particular set of surveillance interval and score requirements. The evaluation result is presented in Fig.~\ref{Graph_Time_Kvisit}. As the result shows with the $k$, the model execution time increases. Since a point under resilient surveillance needs to be visited by $k+1$ UAVs within a specific time frame, a larger $k$ tightens the solution space, thus increasing the searching time.

\noindent\textbf{Impact of the Surveillance Interval.} The solving time of the proposed formal model depends on the surveillance time interval (between two consecutive visits) threshold (maximum). We perform the evaluation varying continuous surveillance and resilient surveillance thresholds. 
The rest of the input, including continuous and resiliency surveillance score requirements, remains the same. The corresponding results are shown in Fig.~\ref{Graph_Time_VisitInterval} and Fig.~\ref{Graph_Time_ResInterval}.
We can observe that when the surveillance interval requirement is relaxed (i.e., with the increase in the threshold), the time to solve the model reduces. 
%

\noindent\textbf{Impact of the Surveillance Period.} The impact of surveillance period (in time steps/units) on the execution time is presented in Fig.~\ref{Graph_Time_Period} and, as the graph shows, the execution time grows rapidly with the period. Each time step is associated with a number of clauses. Hence, if the surveillance time expands, the number of clauses grows, which ultimately increases the model solving time. 

\section{Conclusion}
\label{Sec:Conclusion}

Overhead power transmission lines in a smart grid require regular assessment for reliable and uninterrupted operation, especially considering physical (potential) damages due to natural calamities, aging factors, technical errors, or physical attacks. The emergence of UAV technology provides the opportunity to keep this critical infrastructure under continuous surveillance.
%
In this work, we have proposed a formal framework that synthesizes the trajectory plan as well as the refueling schedules for a given set of UAVs, performing a continuous surveillance of the lines to satisfy various critical line monitoring and resiliency requirements. The resiliency surveillance requirement for a point ensures that if $k$ UAVs fail or are compromised, there is still a UAV to collect the data at the point no later than a threshold time.
%
We have implemented the proposed model and evaluated the tool's ability to analyze surveillance characteristics as well as its scalability. 
%

\bibliographystyle{unsrt}
\bibliography{References}

\appendix

\subsection{LODF Calculation}
\label{App:LODF}

LODFs represent the sensitivity of a power system during a contingency condition such as line outages~\cite{guo2009direct}. During the outage, pre-outage power flowing through the affected line is distributed to the other lines with respect to a sensitivity factor matrix, i.e., LODFs. Each element of the matrix represents if one line is tripped what percent of the power of the affected line will be shared by another line. Thus, LODFs are used to calculate the linear impact of contingencies in power system simulator. Therefore, in a power system during the outage of $l_2$, post contingency power flow of $l_1$ is:
$$P_{l_1}^{post}= P_{l_1}^{pre} + L_{l_1,l_2} * P_{l_2}^{pre}$$
where $P_{l_1}^{pre}$ and $P_{l_2}^{pre}$ are the pre-outage power flows of $l_1$  and $l_2$ respectively and $L_{l_1,l_2}$ is LODF of $l_1$ with respect to the outage of $l_2$. 

To calculate the LODF matrix for a power system~\cite{al2016simulation}, we first need to calculate the $Y$ matrix:
\begin{gather}
Y =\begin{bmatrix}
   y_{11} &  y_{12} & .  & y_{1n}\\  
   y_{21} & y_{22}  & . & y_{2n}\\
   .   & .  & . & .\\
   .  & .  & . & .\\
   y_{n1} & y_{n2} & . & y_{nn}\\
   \end{bmatrix}
\end{gather}
with the help of the following equations:
$$y_{ij}=-\frac{1}{z_{ij}}$$ 
$$ y_{ii}=\sum_{\frac{i=1}{i\ne j}}^{n} y_{ij} $$
Here, $z_{ij}$ is the impedance of the line between bus i and j. 

If we assume bus 1 is the slack bus, eliminating the $1^{st}$ row and $1^{st}$ column from $Y$, we will receive:
\begin{gather}
\hat{Y}=\begin{bmatrix}
   y_{22} & y_{23}  & . & y_{2n}\\
    y_{32} &  y_{33} & .  & y_{1n}\\  
   .   & .  & . & .\\
   .   & .  & . & .\\
   y_{n2} & y_{n3} & . & y_{nn}\\
   \end{bmatrix}
\end{gather}

The sensitivity matrix, $X$, is calculated from $\hat{Y}$:
\begin{gather}
X=\begin{bmatrix}
   0 & 0  \\
   0 & \hat{Y}^{-1} \\
   \end{bmatrix}
\end{gather}
 
 Finally, the LODF of line $l_1$, due to the outage of the line $l_2$, will be calculated by the following equation
 \begin{equation}
     L_{l_1,l_2}=\frac{{\frac{z_{ij}}{z_{kl}}}(X_{il}-X_{jl}-X_{ik}+X_{jk})}{z_{kl}-(X_{ii}+X_{jj}-2X_{ij}}
\end{equation}

\subsection{Criticality Weight Assignment}
\label{App:Weight}

The PI value of a transmission line can be considered directly as the criticality weight. However, loads at the buses frequently changes and so the criticality values. If the changes are minor, it is not worthwhile to modify the surveillance plan. Therefore, it is advantageous to assign qualitative criticality weights ignoring the small variances.
In that case, if $K$ qualitative criticality levels are assumed, one weight assignment approach can be dividing the criticality scale (from the minimum to the maximum) into $K$ number of equally sized ranges. While the first range includes the smallest PI values, the $K$'th range has the highest ones. The lines with the PI values falling within the criticality range $i$ ($1 \le i \le K$) are considered to have the criticality weight $i$. However, better approaches are possible like below where the number of levels ($K$) can be realized according to the distribution of the PI values. We present a mechanism below that provides qualitative weights to the transmission lines, after finding the minimum value of $K$.


\begin{algorithm}[t]
\caption{Criticality-Ranking-Algorithm}
\label{Algo:Criticality-Ranking}
\begin{algorithmic}

\STATE Input: $\mathbb{X} = \{x_1, \ldots, x_i, \ldots, x_N\}$ is the set of data points (PIs), where $N = |\mathbb{L}|$.
\STATE Input: $D$ is the maximum distance (between a data point and the cluster center) allowed within a cluster.
\STATE Input: $K$ is the initial number of ranks (clusters). 

\STATE Output: Let $\mathbb{R} = \{r_1, \ldots, r_i, \ldots, r_N\}$ is the set of criticality weights of the transmission lines.

\STATE Let $\mathbb{C}_1, \ldots, \mathbb{C}_j, \ldots, \mathbb{C}_K$ be the set of clusters to which the data points are to be assigned. Each cluster $\mathbb{C}_j$ has a cluster center (or mean) $c_j$. Let $\mathbb{C} = \{c_1, \ldots, c_j, \ldots, c_K\}$.
\STATE Let parameter $\mathit{Status}$ denote if there is a valid clustering. 
\STATE Initialize: $\mathit{Status}$ := FALSE

\WHILE {TRUE}
	\STATE Clustering-Algorithm($\mathbb{X}$, $K + 1$)
	\STATE $K$ := $K + 1$
	
	\STATE Calculate the distance ($d_{i, j}$) between each data point $x_i$ and each cluster center $c_j$.	
	
	\IF {The maximum of $d_{i, j}$s $> D$}
		\IF {Status = TRUE}
			\STATE Consider the last cluster
			\FOR {each  $\mathbb{C}_j$}
				\FOR {each $x_i \in \mathbb{C}_j$}
					\STATE $r_i$ := $c_j$
				\ENDFOR
			\ENDFOR
			\RETURN $\mathbb{R}$
		\ELSE
			\STATE Clustering-Algorithm($\mathbb{X}$, $K + 1$)
		\ENDIF
	\ELSE
		\STATE {Status := TRUE}
		\STATE Clustering-Algorithm($\mathbb{X}$, $K - 1$)
	\ENDIF
\ENDWHILE
\end{algorithmic}
\end{algorithm}


Algorithm~\ref{Algo:Criticality-Ranking} can do weight assignment based on a clustering method. In particular, the algorithm invokes a procedure named Algorithm~\ref{Algo:Clustering} to form a set of clusters from the PIs following the $K$-means clustering method~\cite{han2011data}. The PI values will be fed to Algorithm~\ref{Algo:Criticality-Ranking} as inputs while the qualitative criticality weights of the transmission lines will be returned. 
The lines with the PI values falling within the same cluster receive the same criticality score, derived from the cluster mean. 
The algorithm finds the minimum $K$ based on a threshold distance value. While the threshold value ensures a particular score is assigned to a set of lines whose PI values are close to each other (no more than a distance from the cluster centroid), the minimization of $K$ reduces the impact of small variances in PI values (which are also not exact measures of the criticality) for prioritizing the lines. 

\begin{algorithm}[t]
\caption{Clustering-Algorithm}
\label{Algo:Clustering}
\begin{algorithmic}

\STATE Input: $\mathbb{X} = \{x_1, \ldots, x_i, \ldots, x_N\}$. 
\STATE Input: $K$, the number of clusters.
\STATE Output: Clusters ($\mathbb{C}_1, \ldots, \mathbb{C}_j, \ldots, \mathbb{C}_K$) and corresponding cluster centers or means ($\mathbb{C}$).
\STATE Initialize: $\mathbb{C}_j$s are initially empty.
\STATE Initialize: $c_j$s are arbitrarily chosen from the data points $x_i$s.

\WHILE {TRUE}
	\STATE Calculate the distance ($d_{i, j}$) between each data point $x_i$ and each cluster center $c_j$.

	\STATE Assign $x_i$ to the cluster ($\mathbb{C}_j$) with center $c_j$ if  $d_{i, j}$ is the minimum.
	
	\IF {No change in the members of $\mathbb{C}_j$s}
		\RETURN $\mathbb{C}$ and $\mathbb{C}$s;
	\ENDIF
	
	\FOR {$c_j \in \mathbb{C}$} 
		\STATE Update cluster center as $c_j = \sum_{x_i \in \mathbb{C}_j} x_i /|\mathbb{C}_j|$
	\ENDFOR		
\ENDWHILE
\end{algorithmic}
\end{algorithm}

\end{document}